\documentclass[a4paper]{jpconf}
\usepackage{graphicx}

\usepackage{amsmath}
\usepackage{graphicx}
\usepackage{bbm}

\newcommand{\nc}{\newcommand}
\nc{\non}{\nonumber}
\nc{\hc}{\hbox {H.c.}} 
\nc{\noi}{\noindent}
\nc{\barx}{\bar{x}}
\nc{\pbarn}{\;\hbox {pb}}
\nc{\fbarn}{\;\hbox {fb}}
\nc{\lsp}{\;\;\;\;\;}
\nc{\Lsp}{\;\;\;\;\;\;\;\;\;\;}  
\nc{\LLsp}{\lspace \lspace}
\nc{\lra}{\longrightarrow}
\nc{\beq}{\begin{equation}}  \nc{\eeq}{\end{equation}}
\nc{\bea}{\begin{eqnarray}}  \nc{\eea}{\end{eqnarray}}
\nc{\baa}{\begin{array}}     \nc{\eaa}{\end{array}}
\nc{\bit}{\begin{itemize}}   \nc{\eit}{\end{itemize}}
\nc{\ben}{\begin{enumerate}} \nc{\een}{\end{enumerate}}
\nc{\bce}{\begin{center}}    \nc{\ece}{\end{center}}
\nc{\bpm}{\begin{pmatrix}}   \nc{\epm}{\end{pmatrix}}
\nc{\bvt}{\begin{verbatim}}  \nc{\evt}{\end{verbatim}}

\def\gesim{\,{\raise-3pt\hbox{$\sim$}}\!\!\!\!\!{\raise2pt\hbox{$>$}}\,}
\def\lesim{\,{\raise-3pt\hbox{$\sim$}}\!\!\!\!\!{\raise2pt\hbox{$<$}}\,}

\def\gev{\;\hbox{GeV}}
\def\tev{\;\hbox{TeV}}

\def\lsp{\qquad}
\def\lsim{\lesim}
\def\gsim{\gesim}
\def\hc{\hbox{H.c.}}
\def\vev{vacuum expectation value}

\def\zBB{{\mathbbm Z}}

\nc{\lam}{\lambda}
\nc{\Lam}{\Lambda}
\nc{\Lams}{\Lambda^2}
\nc{\mws}{m_W^2}
\nc{\mzs}{m_Z^2}
\nc{\mts}{m_t^2}
\nc{\mh}{m_h}
\nc{\mhs}{m_h^2}
\nc{\mvp}{m_\vp}
\nc{\mvps}{m_\vp^2}
\nc{\mw}{m_W}
\nc{\mz}{m_Z}
\nc{\mt}{m_t}
\nc{\mH}{m_{H^\pm}}
\nc{\mA}{m_A}
\nc{\mS}{m_S}
\nc{\vp}{\varphi}
\nc{\mpl}{m_{\rm Pl}}
\nc{\sbb}{s_\beta}
\nc{\cbb}{c_\beta}
\nc{\sba}{s_{\beta-\alpha}}
\nc{\cba}{c_{\beta-\alpha}}
\nc{\stb}{s_{2\beta}}
\nc{\ctb}{c_{2\beta}}
\nc{\mb}{m_b}
\nc{\mbs}{m_b^2}
\nc{\tgb}{\tan\beta}
\nc{\tgbs}{\tan^2\beta}
\nc{\ctbs}{\cot^2\beta}

\nc{\lamp}{\lambda_H}
\nc{\lamvp}{\lambda_\varphi}
\nc{\lamx}{\lambda_x}
\nc{\xf}{x_f}

\nc{\co}{{\bf ??? }}

\renewcommand{\Re}{{\rm Re\thinspace}}
\renewcommand{\Im}{{\rm Im\thinspace}}

\begin{document}
\title{Tuned Two-Higgs-Doublet Model}

\author{B. Grzadkowski$^1$, P. Osland$^2$}

\address{$^1$ Institute of Theoretical Physics, 	Faculty of Physics, University of Warsaw, 
Ho\.za 69, PL-00-681 Warsaw, Poland}
\address{$^2$ Department of Physics and Technology, University of Bergen,
Postboks 7803, N-5020 Bergen, Norway}
\ead{$^1$ Bohdan.Grzadkowski@fuw.edu.pl}
\ead{$^2$ per.osland@ift.uib.no}

\begin{abstract}
  We consider a Two-Higgs-Doublet Model (2HDM) constrained by the 
  condition that assures
  cancellation of quadratic divergences up to the leading two-loop order.
  Regions in the parameter space consistent with existing experimental constraints
  and with the cancellation  condition are determined. The possibility for CP violation in
  the scalar potential is discussed and regions of $\tgb-M_{H^\pm}$
  with substantial amount of CP violation are found. The model allows to ameliorate
  the little hierarchy problem by lifting the minimal scalar Higgs
  boson mass and by suppressing the quadratic corrections to
  scalar masses. The cutoff originating from the naturality arguments
  is therefore lifted from $\sim 0.6\tev$ in the Standard Model 
  to $\gsim 2.5 \tev$ in the 2HDM, depending on the mass of the
  lightest scalar.
\end{abstract}

\section{Introduction}
This project aims at extending the Standard Model (SM) in such a way that 
there would be no quadratic divergences 
up to the leading order at the two-loop level of the perturbation expansion.  
The quadratic divergences were first discussed within the SM by
Veltman~\cite{Veltman:1980mj}, who, adopting dimensional
reduction~\cite{Siegel:1979wq}, found the following quadratically divergent
one-loop contribution to the Higgs boson ($h$) mass 
\begin{equation}
\delta^{\rm (SM)} \mhs = \frac{\Lams}{\pi^2 v^2}\left[\frac32
\mts-\frac18\left(6\mws+3\mzs\right) - \frac38 \mhs \right],
\label{hcor}
\end{equation}
where $ \Lam$ is a UV cutoff 
and $v \simeq 246 \gev $ denotes the \vev\ of the scalar doublet. The
issue of quadratic divergences was then investigated further in ~\cite{Osland:1992ay} and 
\cite{Einhorn:1992um}.

Within the SM precision measurements require a light Higgs boson, therefore the
correction (\ref{hcor}) exceeds the mass itself even for small values
of $ \Lam $, e.g. for $\mh = 130 \gev$ one obtains $\delta^{\rm (SM)}
\mh^2 \simeq \mh^2$ already for $\Lam \simeq 600 \gev$. On the other
hand, if we assume that the scale of new physics is widely separated
from the electro-weak scale, then constraints that emerge from
analysis of operators of dimension 6 require $\Lam \gsim$ a few TeV.
The lesson from this observation is that regardless of what physics lies beyond the SM, 
some amount of fine tuning is necessary; either we tune to
lift the cutoff above $\Lam \simeq 600 \gev$, or we tune when
precision observables measured at LEP are fitted.
Tuning both in corrections to the Higgs mass and in LEP physics is, of
course, also a viable alternative which we are going to explore below.
So, we will look for new physics in the TeV range which will allow to
lift the cutoff implied by quadratic corrections to $\mhs$ to the
multi-TeV range {\it and} which will be consistent with all the
experimental constraints---both require some amount of tuning.  Note
that within the SM the requirement $\delta^{\rm (SM)} \mhs = 0$
implies an unrealistic value of the Higgs boson mass $\mh \simeq
310~\gev$.

Here we are going to argue that the
Two-Higgs-Doublet Model (2HDM) in certain region of its parameter space can 
soften the little hierarchy problem both by suppressing quadratic corrections
to scalar masses {\it  and} it allows to lift the central value for the 
lightest Higgs mass.

\section{ The Two-Higgs-Doublet Model}
\label{non-IDM}

In order to accommodate CP violation we consider here a 2HDM
with softly broken $\zBB_2$ symmetry which acts as $\Phi_1\to -\Phi_1$ and $u_R\to -u_R$ (all other fields are neutral). The scalar potential then reads
\begin{eqnarray}
V(\phi_1,\phi_2) &=&  -\frac12 \left\{m_{11}^2\phi_1^\dagger\phi_1 
+ m_{22}^2\phi_2^\dagger\phi_2 + \left[m_{12}^2 \phi_1^\dagger\phi_2 
+ \hc \right]\right\}
 + \frac12 \lam_1 (\phi_1^\dagger\phi_1)^2 
+ \frac12 \lam_2 (\phi_2^\dagger\phi_2)^2 
\nonumber  \\
&& 
+ \lambda_3(\phi_1^\dagger\phi_1)(\phi_2^\dagger\phi_2) 
+ \lambda_4(\phi_1^\dagger\phi_2)(\phi_2^\dagger\phi_1) 
+ \frac12\left[\lambda_5(\phi_1^\dagger\phi_2)^2 + \hc\right] 
\label{2HDMpot}
\end{eqnarray}
The minimization conditions at $\langle \phi_1^0 \rangle = v_1/\sqrt{2}$ and 
$\langle \phi_2^0 \rangle = v_2/\sqrt{2}$ can be formulated as follows:
\beq
m_{11}^2= v_1^2\lam_1+v_2^2(\lambda_{345}-2\nu), \lsp
m_{22}^2=v_2^2\lam_2+v_1^2(\lambda_{345}-2\nu),
\label{min}
\eeq
where $\lambda_{345}\equiv \lam_3+\lam_4+\Re\lam_5$ and 
$\nu\equiv \Re m_{12}^2/(2v_1v_2)$.

\subsection{Quadratic divergences}
\label{one-loop}

At the one-loop level the cancellation of quadratic divergences for the scalar Green's functions 
at zero external momenta ($\Gamma_i$, $i=1,2$) in the 2HDM 
type II model implies~\cite{Newton:1993xc} 
\begin{eqnarray}
\Gamma_1\equiv \frac32 \mw^2 + \frac34 \mz^2 
+ \frac{v^2}{2}\left( \frac32 \lam_1 + \lam_3 + \frac12 \lam_4 \right) 
- 3 \frac{\mb^2}{\cbb^2} = 0,
\label{qdcon1_mod2}\\
\Gamma_2\equiv\frac32 \mw^2 + \frac34 \mz^2 
+ \frac{v^2}{2}\left( \frac32 \lam_2 + \lam_3 + \frac12 \lam_4 \right) 
-3 \frac{\mt^2}{\sbb^2} = 0,
\label{qdcon2_mod2}
\end{eqnarray} 
where $v^2\equiv v_1^2+v_2^2$, $\tan\beta\equiv v_2/v_1$ and we use the
notation: $s_\theta \equiv \sin\theta$ and $c_\theta\equiv \cos\theta$. 
Note that when $\tan\beta$ is large, the two quark contributions can be
comparable. In the type II model
the mixed, $\phi_1-\phi_2$, Green's function is not quadratically divergent.

The quartic couplings $\lambda_i$ can be
expressed in terms of the mass parameters and elements of the rotation matrix 
needed for diagonalization of the scalar masses (see, for
example, Eqs.~(3.1)--(3.5) of \cite{ElKaffas:2007rq}). 
Therefore, for a given choice of $\alpha_i$'s, the
squared neutral-Higgs masses $M_{1}^2$, $M_{2}^2$ and $M_3^2$ can be
determined from the cancellation conditions
(\ref{qdcon1_mod2})--(\ref{qdcon2_mod2}) in terms of $\tgb$, $\mu^2$
and $M_{H^\pm}^2$.
It is worth noticing that scalar masses resulting from a scan over $\alpha_i$, $M_{H^\pm}$ and 
$\tgb$ exhibit a striking mass
degeneracy in the case of
large $\tan\beta$: $M_1 \simeq M_2 \simeq M_3
\simeq \mu^2+4\mb^2$. 

At the two-loop level the leading contributions to quadratic divergences are
of the form of $ \Lam^2 \ln \Lam$. They could be determined adopting a method 
noticed by Einhorn and Jones~\cite{Einhorn:1992um}, so that the cancellation 
conditions for quadratic divergences up to the leading two-loop order read:
\beq
\Gamma_1+\delta \Gamma_1=0 \lsp {\rm and} \lsp \Gamma_2+\delta \Gamma_2=0 
\label{2-loop-con}
\eeq
with
\bea
\delta \Gamma_1 &=& \frac{v^2}{8} [
9 g_2 \beta_{g_2} + 3 g_1 \beta_{g_1} + 6\beta_{\lambda_1} + 4 \beta_{\lambda_3} + 2 \beta_{\lambda_4}]\ln\left(\frac{\Lambda}{\bar\mu}\right)\\
\delta \Gamma_2 &=& \frac{v^2}{8} [
9 g_2 \beta_{g_2} + 3 g_1 \beta_{g_1} + 6\beta_{\lambda_2} + 4 \beta_{\lambda_3} + 2 \beta_{\lambda_4}
-24 g_t \beta_{g_t}]\ln\left(\frac{\Lambda}{\bar\mu}\right)
\eea
where $\beta$'s are the appropriate beta functions while $\bar \mu$ is the renormalization scale.
Hereafter we will be solving the conditions (\ref{2-loop-con}) for the scalar masses $M_i^2$ for a given set of $\alpha_i$'s, $\tgb$, $\mu^2$ and $M_{H^\pm}^2$. For the renormalization scale
we will adopt $v$, so $\bar\mu=v$. Then those masses together with the corresponding coupling constants, will be used to find predictions of the model for various observables which then can be checked against experimental data.

\begin{figure}[t]
\centering
\includegraphics[width=5.cm]{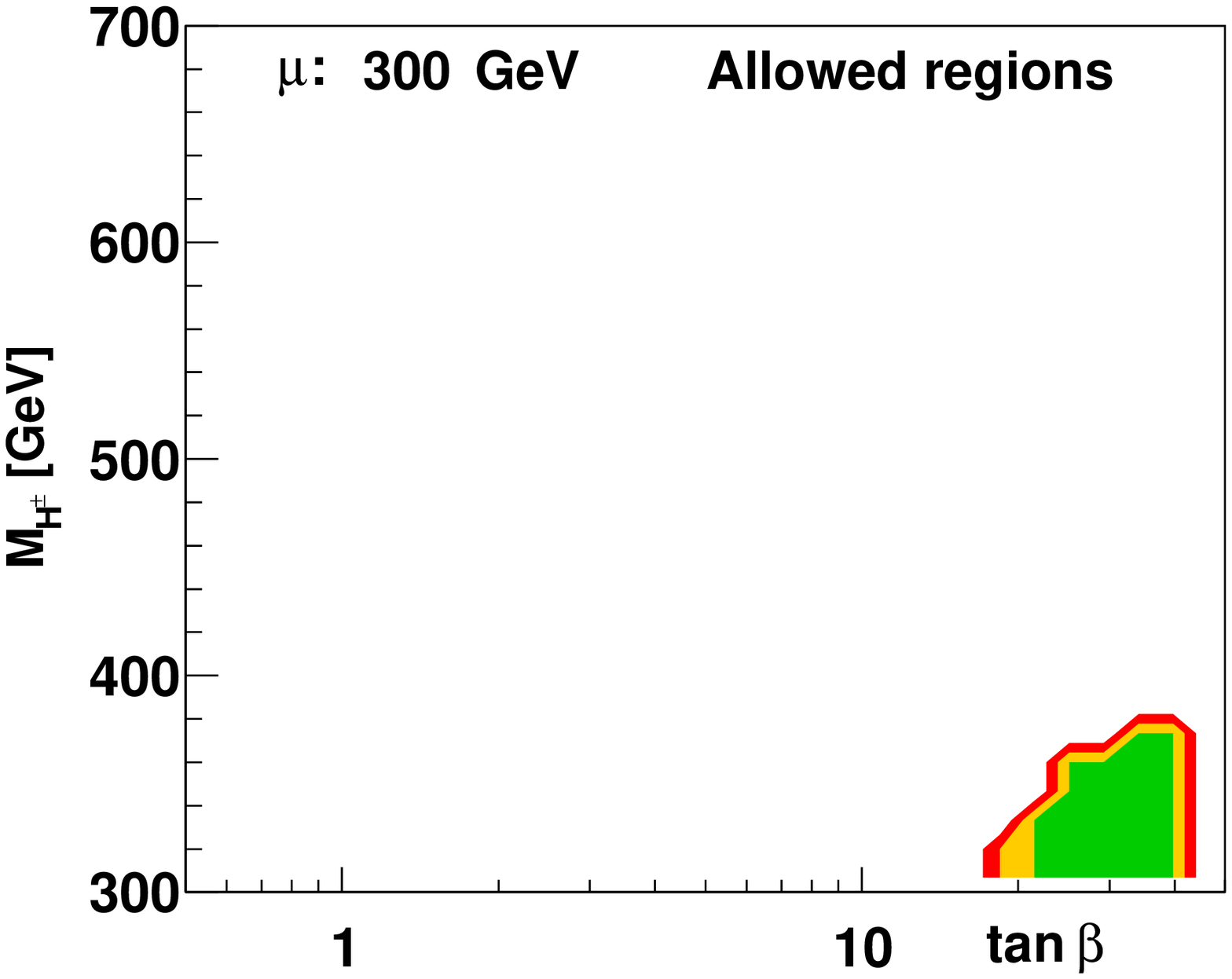}
\includegraphics[width=5.cm]{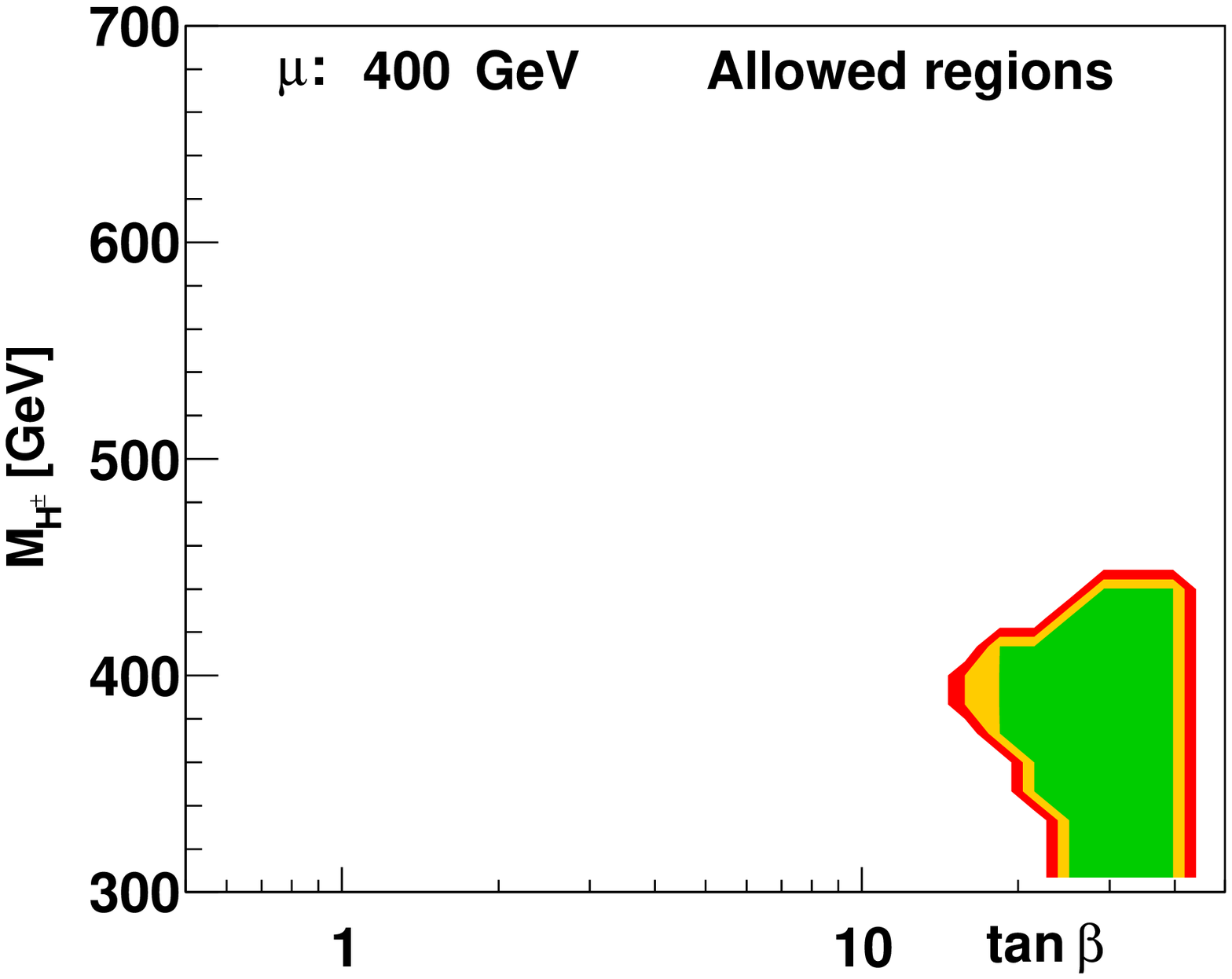}
\includegraphics[width=5.cm]{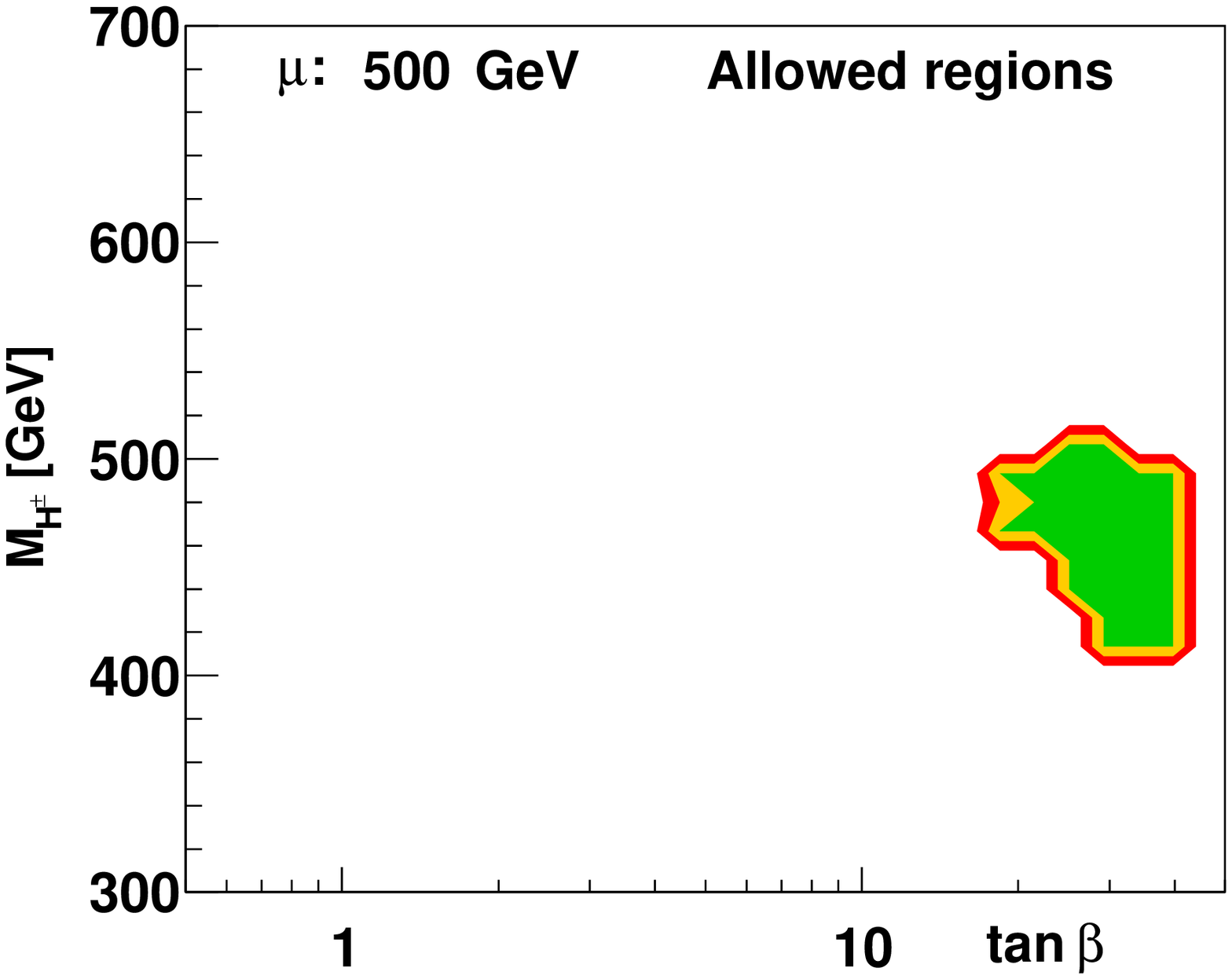}
\caption{\label{Fig:allowed-2500-300-400-500}
  Two-loop allowed regions in
  the $\tan\beta$--$M_{H^\pm}$ plane, for $\Lam=2.5\tev$, for $\mu=300, 400, 500\gev$ 
  (as indicated).  Red: positivity
  is satisfied; yellow: positivity and unitarity both satisfied;
  green: also experimental constraints satisfied at the 95\% C.L., as
  specified in the text. }
\end{figure}

\subsection{Allowed regions}
\label{sec:allowed}

In order to find phenomenologically acceptable regions in the 
parameter space we impose the following experimental constraints:
the oblique parameters $T$ and $S$, $B_0-\bar{B}_0$ mixing, $B\to X_s \gamma$,
$B\to \tau \bar\nu_\tau X$, $B\to D\tau \bar\nu_\tau$, LEP2 Higgs-boson non-discovery,
$R_b$, the muon anomalous magnetic moment and the electron electric dipole moment
(for details concerning the experimental constraints, see
refs.~\cite{Grzadkowski:2009bt,ElKaffas:2007rq,WahabElKaffas:2007xd}).
Subject to all these constraints, we find allowed solutions of (\ref{2-loop-con}). 
For instance, imposing all the experimental constraints 
we find allowed regions in the $\tan\beta$--$M_{H^\pm}$ plane as 
illustrated by the red domains in the $\tan\beta$--$M_{H^\pm}$ plane, see
Fig.~\ref{Fig:allowed-2500-300-400-500} 
for fixed values of $\mu$. The allowed regions
were obtained scanning over the mixing angles $\alpha_i$ and solving
the two-loop cancellation conditions (\ref{2-loop-con}). Imposing
also unitarity in the Higgs-Higgs-scattering sector 
\cite{Kanemura:1993hm,Akeroyd:2000wc,Ginzburg:2003fe}
(yellow regions), the allowed regions are only slightly
reduced. Requiring that also experimental constraints 
are satisfied the green regions are obtained. 

\begin{figure}[ht]
\centering
\includegraphics[width=12cm]{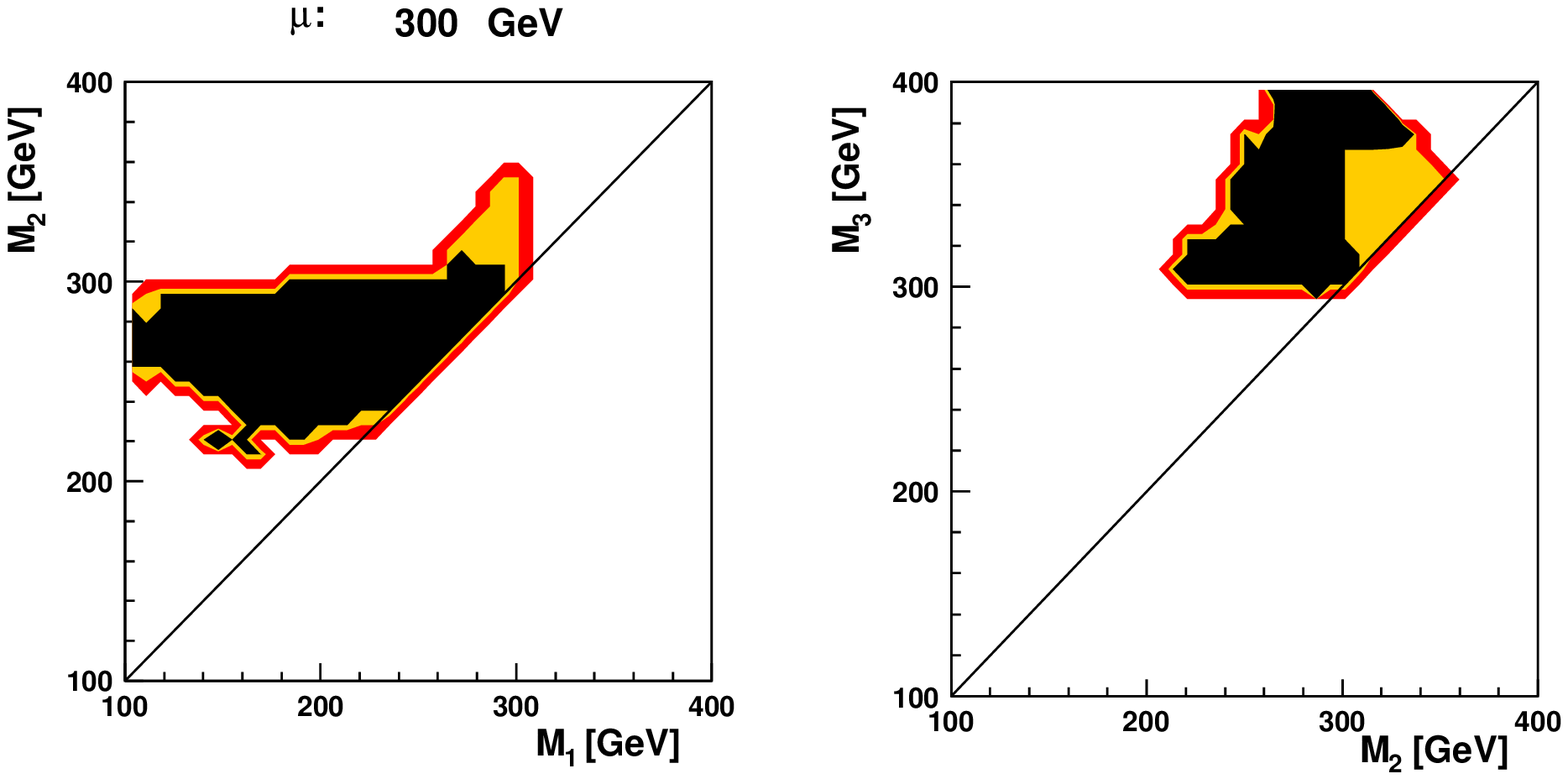}
\includegraphics[width=12cm]{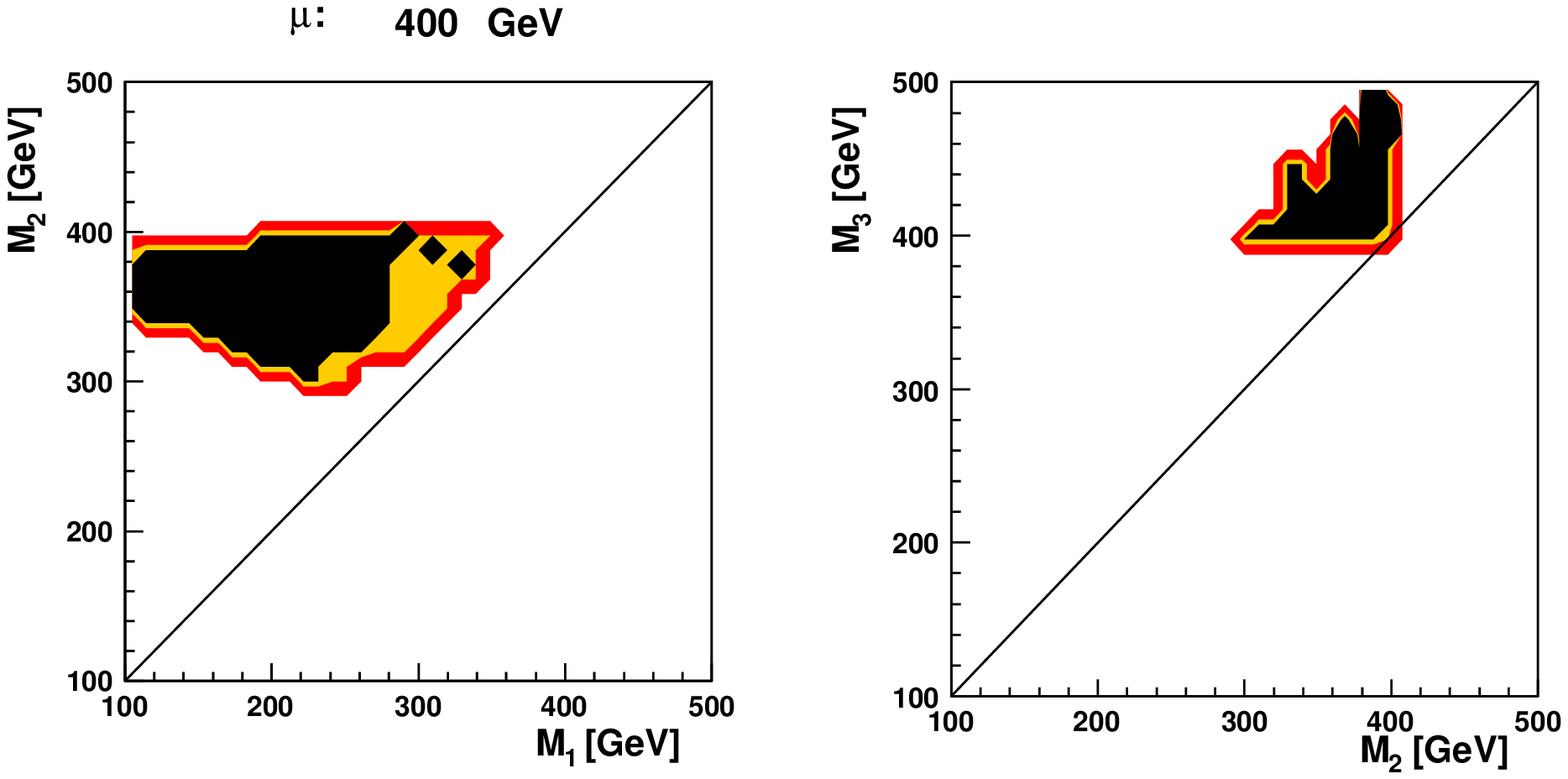}
\includegraphics[width=12cm]{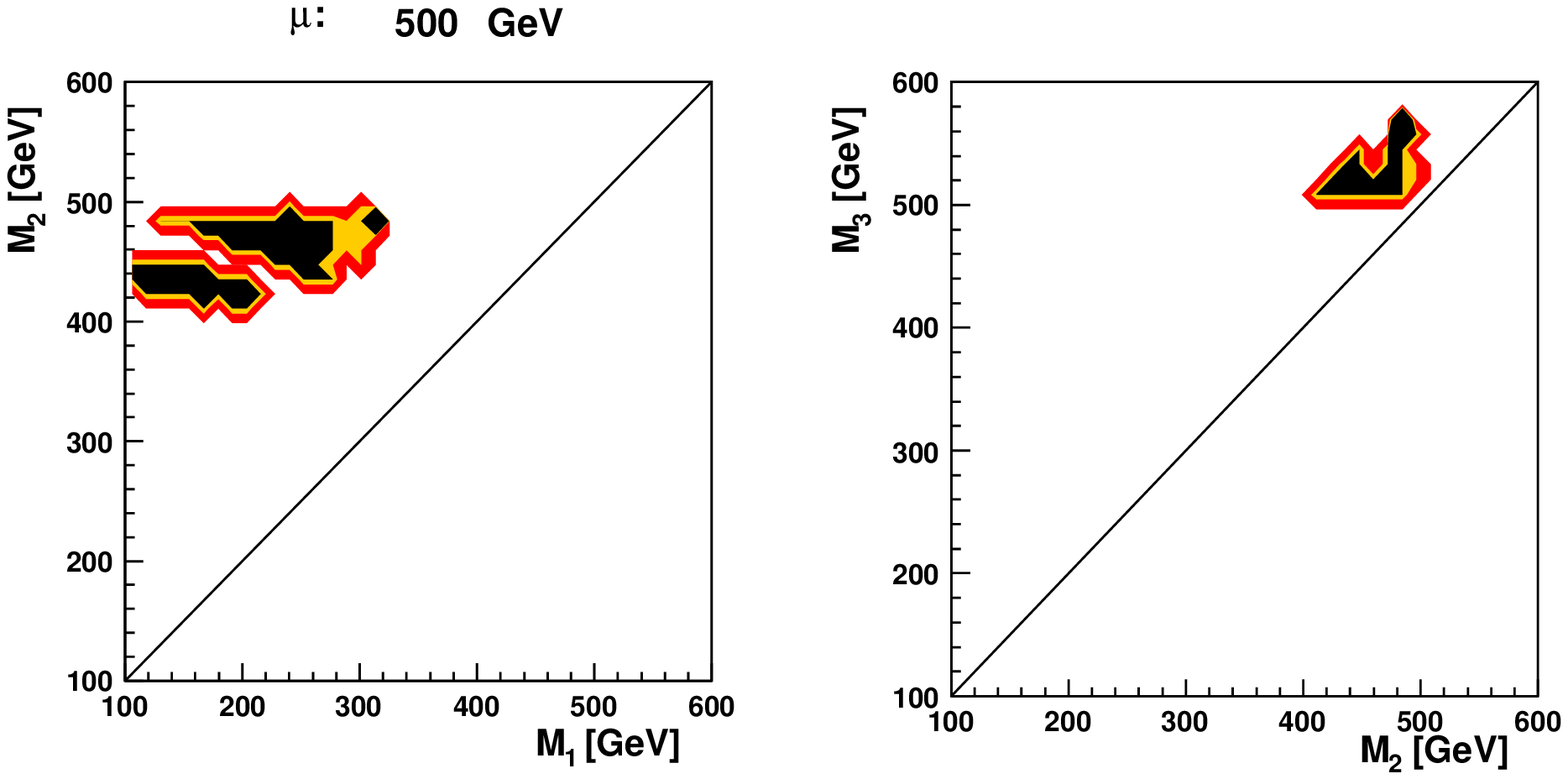}
\caption{
  Two-loop distributions of allowed masses $M_2$ vs $M_1$ (left panels)
  and $M_3$ vs $M_2$ (right) for $\Lam=2.5\tev$, resulting from a scan over the full
  range of $\alpha_i$, $\tan\beta \in (0.5,50)$ and $M_{H^\pm} \in
  (300,700)\gev$, for $\mu=300, 400, 500~{\rm GeV}$. Red: Positivity
  is satisfied; yellow: positivity and unitarity both satisfied;
  green: also experimental constraints satisfied at the 95\% C.L., as
  specified in the text. }
\label{Fig:2-loop-masses-2500}
\end{figure}

For parameters that are consistent with unitarity, positivity, experimental
constraints and the two-loop cancellation conditions (\ref{2-loop-con}), we show in Fig.~\ref{Fig:2-loop-masses-2500} scalar masses resulting from a scan over $\alpha_i$, 
$M_{H^\pm}$ and $\tgb$. 
As we have noticed for the one-loop spectrum, large $\tan\beta$ implies
similar scalar masses. This is indeed what is being observed in Fig.~\ref{Fig:2-loop-masses-2500}
also for the two-loop case. The allowed solutions ``peak'' around $M_{H^\pm}\sim \mu$ with $20 \lsim \tan\beta \lsim 50$. 

\subsection{CP violation}
\label{cpv}

Here we will verify the
possibility of having CP violation in the scalar potential
(\ref{2HDMpot}), subject to the two-loop cancellation of quadratic
divergences (\ref{2-loop-con}).  In
order to parametrize the magnitude of CP violation we adopt the 
$U(2)$-invariants introduced by Lavoura and Silva \cite{Lavoura:1994fv}
(see also \cite{Branco:2005em}). However here we 
use the basis-invariant
formulation of these invariants $J_1$, $J_2$ and $J_3$ as
proposed by Gunion and Haber \cite{Gunion:2005ja}.
As is proven there (theorem \#4) the Higgs sector is
CP-conserving if and only if $\Im J_i=0$ for all $i$. In the basis adopted
here the invariants read~\cite{Grzadkowski:2009bt}:
\begin{eqnarray}
\Im J_1&=&-\frac{v_1^2v_2^2}{v^4}(\lambda_1-\lambda_2)\Im \lambda_5,
\label{Eq:ImJ_1} \\
\Im J_2&=&-\frac{v_1^2v_2^2}{v^8}
\left[\left((\lambda_1-\lambda_3-\lambda_4)^2-|\lambda_5|^2\right) v_1^4
+2(\lambda_1-\lambda_2) \Re \lambda_5 v_1^2v_2^2\right.\nonumber\\
&&\hspace*{1.2cm}\left.
-\left((\lambda_2-\lambda_3-\lambda_4)^2-|\lambda_5|^2\right) v_2^4\right]
\Im \lambda_5,\label{Eq:ImJ_2} \\
\Im J_3&=&\frac{v_1^2v_2^2}{v^4}(\lambda_1-\lambda_2)
(\lambda_1+\lambda_2+2\lambda_4)\Im \lambda_5.
\label{Eq:ImJ_3}
\end{eqnarray}
It is seen that there is no CP violation when $\Im\lambda_5=0$,
see \cite{Grzadkowski:2009bt} for more details.

\begin{figure}[ht]
\centering
\includegraphics[width=16cm]{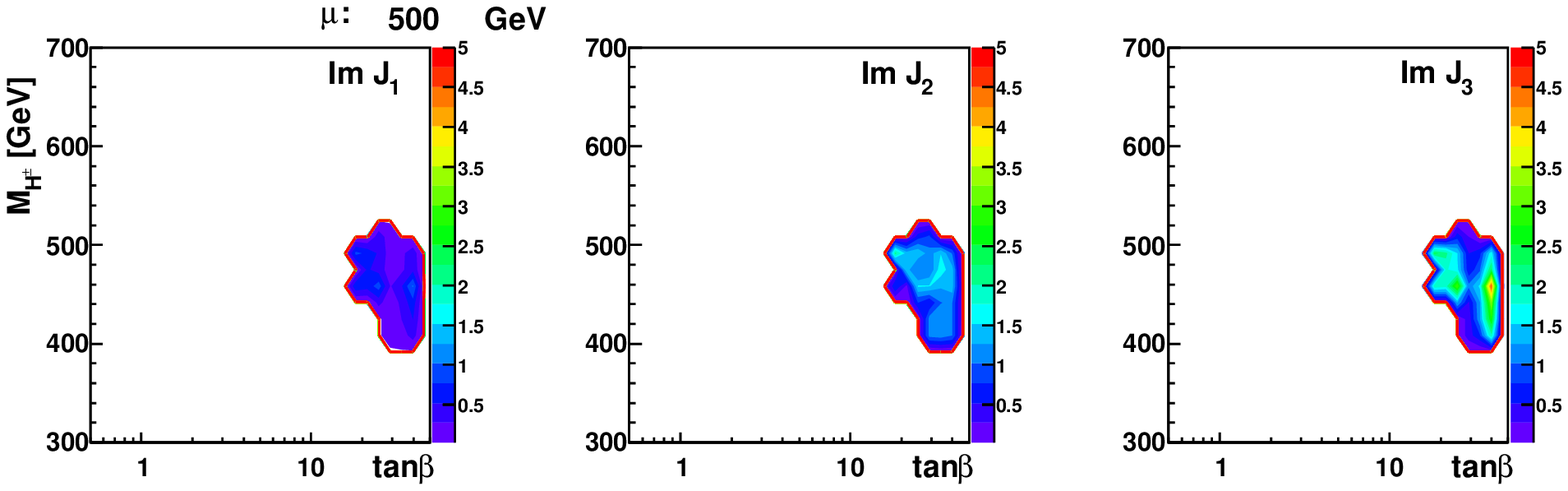}
\includegraphics[width=16cm]{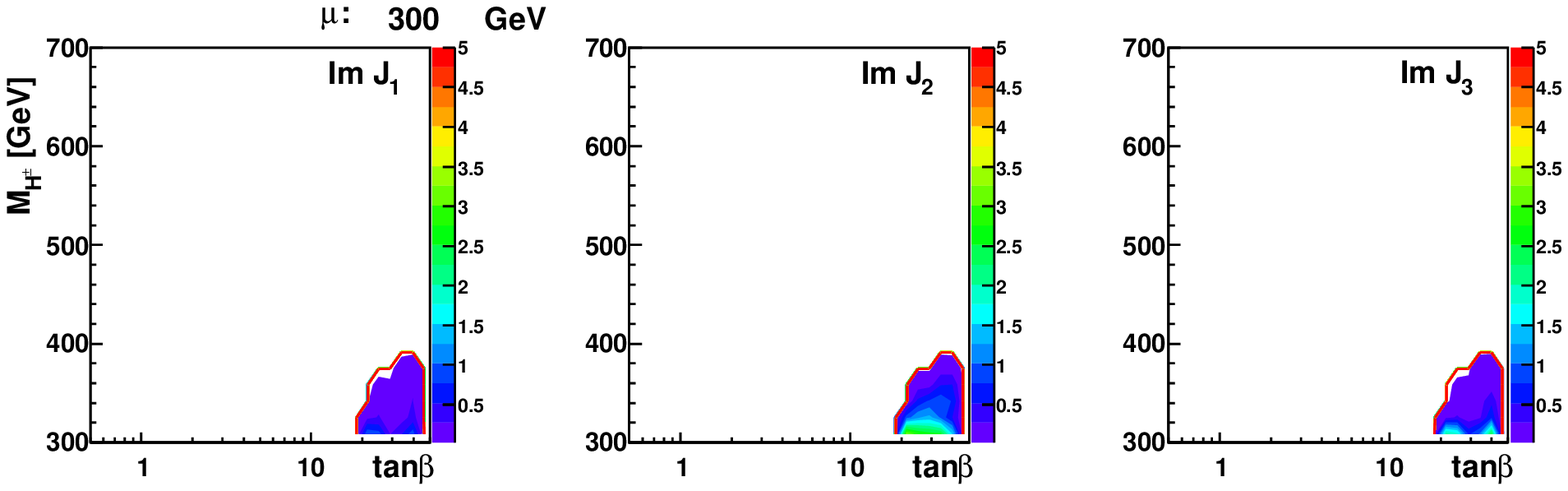}
\caption{
  Absolute values of the imaginary parts of the  $U(2)$-invariants $|\Im J_i|$ at the
  two-loop level for $\Lam=2.5\tev$, for
  $\mu=500~{\rm GeV}$ (top) and $\mu=300~{\rm GeV}$ (bottom). 
  The color coding in units $10^{-3}$ is given along the
  right vertical axis.}
\label{Fig:imj-2500-300-500}
\end{figure}

As we have noted earlier, $\tgb$ above $\sim 40$ implies approximate
degeneracy of scalar masses. That could jeopardize the CP
violation in the potential since it is well known that the exact degeneracy $M_1=M_2=M_3$
results in vanishing invariants $\Im J_i$ and no CP violation (exact
degeneracy implies $\Im \lambda_5=0$). Using the one-loop conditions
(\ref{qdcon1_mod2})--(\ref{qdcon2_mod2}) one immediately finds that
$\lambda_1-\lambda_2=4(\mb^2/\cbb^2-\mt^2/\sbb^2)/v^2$, which implies
\begin{equation}
\Im J_1 = 4\, \Im \lambda_5 (\cbb^2\mt^2-\sbb^2 \mb^2)/v^2
=-4\, \Im \lambda_5 \left(\mb/v\right)^2 + 
{\cal{O}}\left(\Im \lambda_5/\tgb^2\right)
\label{imj1}
\end{equation}
Note that if
$\tgb$ is large then $\Im J_1$ is suppressed not only by $\Im
\lambda_5 \simeq 0$ (as caused by $M_1\simeq M_2\simeq M_3$) but also
by the factor $(\mb^2/v^2)$, as implied by the cancellation conditions
(\ref{qdcon1_mod2})--(\ref{qdcon2_mod2}). The same suppression factor
appears for $\Im J_3$. The case of $\Im J_2$ is more involved, however
when $\mb^2/v^2$ is neglected all the invariants
(\ref{Eq:ImJ_1})--(\ref{Eq:ImJ_3}) have the same simple asymptotic
behavior for large $\tgb$:
\begin{equation}
\Im J_i \sim \Im \lambda_5/\tgbs
\end{equation}
Those conclusions qualitatively remain also
at the two-loop level. For a quantitative illustration
we plot in Fig.~\ref{Fig:imj-2500-300-500} maximal values of the invariants in
the $\tan\beta$--$M_{H^\pm}$ plane with all the necessary constraints
imposed, looking for regions which still allow for substantial CP
violation. At high values of $\tan\beta$ these invariants are of the
order of $10^{-3}$, in qualitative agreement with the discussion
above. It is worth noticing 
that the corresponding invariant in the SM; $\Im Q = \Im (V_{ud} V_{cb} V_{ub}^\star V_{cd}^\star)$~\cite{Bernreuther:2002uj} is of the order of 
$\sim 2\times 10^{-5} \sin \delta_{KM}$ ($V_{ij}$ and $\delta_{KM}$ are elements of the CKM matrix and CP-violating phase, respectively). Therefore the model considered here allows for CP violation at least two orders of magnitude larger than in the SM.

\section{Summary}
\label{sum}
The goal of this project was to build a minimal realistic model which
would ameliorate the little hierarchy problem through
suppression of the quadratic divergences in scalar boson mass
corrections and through lifting the mass of the lightest Higgs
boson. It has been shown that it could be accomplished within the Two-Higgs-Doublet
Model type II. Phenomenological consequences of requiring no
quadratic divergences in corrections to scalar masses 
were discussed.  The 2HDM type II was
analyzed taking into account the relevant existing experimental
constraints. Allowed regions in the parameter space were
determined. An interesting scalar mass degeneracy was noticed for
$\tgb \gsim 40$. The issue of possible CP violation in the scalar
potential was discussed and regions of $\tgb-M_{H^\pm}$ with
substantial strength of CP violation were identified. 
The
cutoff implied by the naturality arguments is lifted from $\sim
600\gev$ in the SM up to at least $\gsim 2.5 \tev$, depending on the mass of
the lightest scalar.
In order to
accommodate a possibility for dark matter a scalar gauge singlet should be
added to the model. 

\ack{B.G. thanks the Organizers of  
PASCOS 2010, the 16th International Symposium on Particles, Strings and Cosmology,
for their warm hospitality during the meeting.
This work is supported in part by the Ministry of Science and Higher
Education (Poland) as research project N~N202~006334 (2008-11). 
B.~G. acknowledges support of the
European Community within the Marie Curie Research \& Training
Networks: ``HEPTOOLS" (MRTN-CT-2006-035505) and ``UniverseNet"
(MRTN-CT-2006-035863).
The research of P.~O. has been supported by the Research Council of
Norway.}

\end{document}